\documentclass[aps,prl,twocolumn,superscriptaddress]{revtex4-1}
\usepackage{graphicx}
\usepackage{dcolumn}
\usepackage{bm}
\usepackage{color}
\usepackage{eso-pic}
\usepackage{amsmath}
\usepackage{xspace}
\newcommand{\LTO}{$\mathrm{LaTiO}_3$\xspace}
\newcommand{\STO}{$\mathrm{SrTiO}_3$\xspace}
\newcommand{\LAO}{$\mathrm{LaAlO}_3$\xspace}

\newcommand{\SN}{$\mathrm{Si}_3\mathrm{N}_4$\xspace}

\usepackage{soul}

\begin{document}

\title{\Large Origin of the dome-shaped superconducting phase diagram in \STO-based interfaces}
\author{A. Jouan}
\affiliation{Laboratoire de Physique et d'\'{E}tude des Mat\'eriaux, ESPCI Paris, Universit\'e PSL, CNRS, Sorbonne Universit\'e, Paris, France.}

\author{S. Hurand}
\affiliation{Laboratoire de Physique et d'\'{E}tude des Mat\'eriaux, ESPCI Paris, Universit\'e PSL, CNRS, Sorbonne Universit\'e, Paris, France.}
\affiliation{Institut Pprime, UPR 3346 CNRS, Universit\'e de Poitiers, ISAE-ENSMA, BP 30179, 86962 Futuroscope-Chasseneuil Cedex, France.}

\author{G. Singh}
\affiliation{Laboratoire de Physique et d'\'{E}tude des Mat\'eriaux, ESPCI Paris, Universit\'e PSL, CNRS, Sorbonne Universit\'e, Paris, France.}
\affiliation{Quantum Device Physics Laboratory, Department of Microtechnology and Nanoscience MC2, Chalmers University of Technology, Gothenburg, Sweden.}

\author{E. Lesne}
\affiliation{Kavli Institute of Nanoscience, Delft University of Technology, Delft, the Netherlands.}
\affiliation{Unit\'e Mixte de Physique CNRS-Thales, Universit\'e Paris-Sud, Universit\'e Paris-Saclay, 1 Av. A. Fresnel, 91767 Palaiseau, France.}

\author{A. Barth\'el\'emy} 
\affiliation{Unit\'e Mixte de Physique CNRS-Thales, Universit\'e Paris-Sud, Universit\'e Paris-Saclay, 1 Av. A. Fresnel, 91767 Palaiseau, France.}

\author{M. Bibes}
\affiliation{Unit\'e Mixte de Physique CNRS-Thales, Universit\'e Paris-Sud, Universit\'e Paris-Saclay, 1 Av. A. Fresnel, 91767 Palaiseau, France.}

\author{C. Ulysse}
\affiliation{Centre for Nanoscience and Nanotechnology, CNRS, Universit\'e Paris-Sud, Universit\'e Paris-Saclay, Boulevard Thomas Gobert, Palaiseau, France.}

\author{G. Saiz}
\affiliation{Laboratoire de Physique et d'\'{E}tude des Mat\'eriaux, ESPCI Paris, Universit\'e PSL, CNRS, Sorbonne Universit\'e, Paris, France.}

\author{C. Feuillet-Palma}
\affiliation{Laboratoire de Physique et d'\'{E}tude des Mat\'eriaux, ESPCI Paris, Universit\'e PSL, CNRS, Sorbonne Universit\'e, Paris, France.}

\author{J. Lesueur}
\affiliation{Laboratoire de Physique et d'\'{E}tude des Mat\'eriaux, ESPCI Paris, Universit\'e PSL, CNRS, Sorbonne Universit\'e, Paris, France.} 

\author{N. Bergeal}
\affiliation{Laboratoire de Physique et d'\'{E}tude des Mat\'eriaux, ESPCI Paris, Universit\'e PSL, CNRS, Sorbonne Universit\'e, Paris, France.}
\date{\today}

\begin{abstract}
A dome-shaped phase diagram of superconducting critical temperature upon doping is often considered as a hallmark of unconventional superconductors. This behavior, observed in two-dimensional electron gases in \STO-based interfaces whose electronic density is controlled by field effect, has not been explained unambiguously yet.  Here, we elaborate a generic scenario for the superconducting phase diagram of  these oxide interfaces based on Schr\"odinger-Poisson numerical simulations of the quantum well and transport experiments on a double-gate field-effect device. We propose that the optimal doping point of maximum $T_c$ marks the transition between a single-band and a fragile two-gap s$_\pm$-wave superconducting state involving  $t_{2g}$ bands of different orbital character. At the optimal doping point, we predict and observe experimentally a bifurcation in the dependence of $T_c$ on the carrier density, which is controlled by the details of the doping execution. Where applying a back-gate voltage triggers the filling of a high-energy $d_\mathrm{xy}$ subband and initiates the overdoped regime, doping with a top-gate delays the filling of the subband and maintains the 2-DEG in the single-band superconducting state of higher $T_c$.
\end{abstract}

\maketitle

The superconducting two-dimensional electron gas (2-DEG) that forms at the interface between two insulating oxides, such as in \LAO/\STO  and \LTO/\STO heterostructures, exhibits  a complex phase diagram controlled by the electron density \cite{caviglia,biscaras2}. While the system is in a weakly insulating state at low density, superconductivity emerges when electrons are added by means of electrostatic gating resorting to a back-gate, a side-gate, or a top-gate geometry \cite{caviglia,hurand,stornaiuoloPRB} (Fig. 1). When the carrier density ($n_\mathrm{2D}$)  increases, the superconducting $T_c$  rises to a maximum value, $T_c^\mathrm{max}$ $\simeq$ 300 mK, before decreasing as doping is further increased. The resulting dome-shaped superconducting phase diagram resembles that observed in other families of superconductors, including high-T$_c$ cuprates, Fe-based superconductors, heavy fermions, and organic superconductors \cite{taillefer,keimer}. Two noticeable doping points are universally observed in the phase diagram of oxide interfaces :  a quantum critical point (QCP) at low density that separates a weakly insulating region and a superconducting one, and a maximum critical temperature point ($T_c^\mathrm{max}$) at an optimal doping.  Despite much research efforts, there is not yet a consensus on the origin of these two points. In \LAO/\STO heterostructures, the interfacial quantum well accommodates a set of discrete $t_{2g}$-based subbands (see insets in Fig. 1) \cite{popovic,delugas,scopigno,salluzzo}.  The $d_\mathrm{xy}$ subbands are energetically the lowest lying orbitals with a pronounced 2D character. Sitting higher in energy in the quantum well, the degenerate $d_\mathrm{xz/yz}$ subbands delocalize  deeper in the substrate, where they recover bulk-like properties, including a high dielectric permittivity and reduced scattering. Several studies have pointed out the connection between superconductivity and the filling threshold of the degenerate $d_\mathrm{xz/yz}$ subbands, whose high density of states favors the emergence of superconductivity \cite{valentinis,singh}. However, the $T_c^\mathrm{max}$ point at the top of the dome, remains largely unexplained. Among a few different scenarios, it has been suggested that  the suppression of $T_c$ in the overdoped regime could result from a strong pair breaking scattering in the presence of opposite-sign gaps s$_\pm$-wave superconductivity \cite{trevisan}.

\begin{figure}[t]
\includegraphics[width=8.5cm]{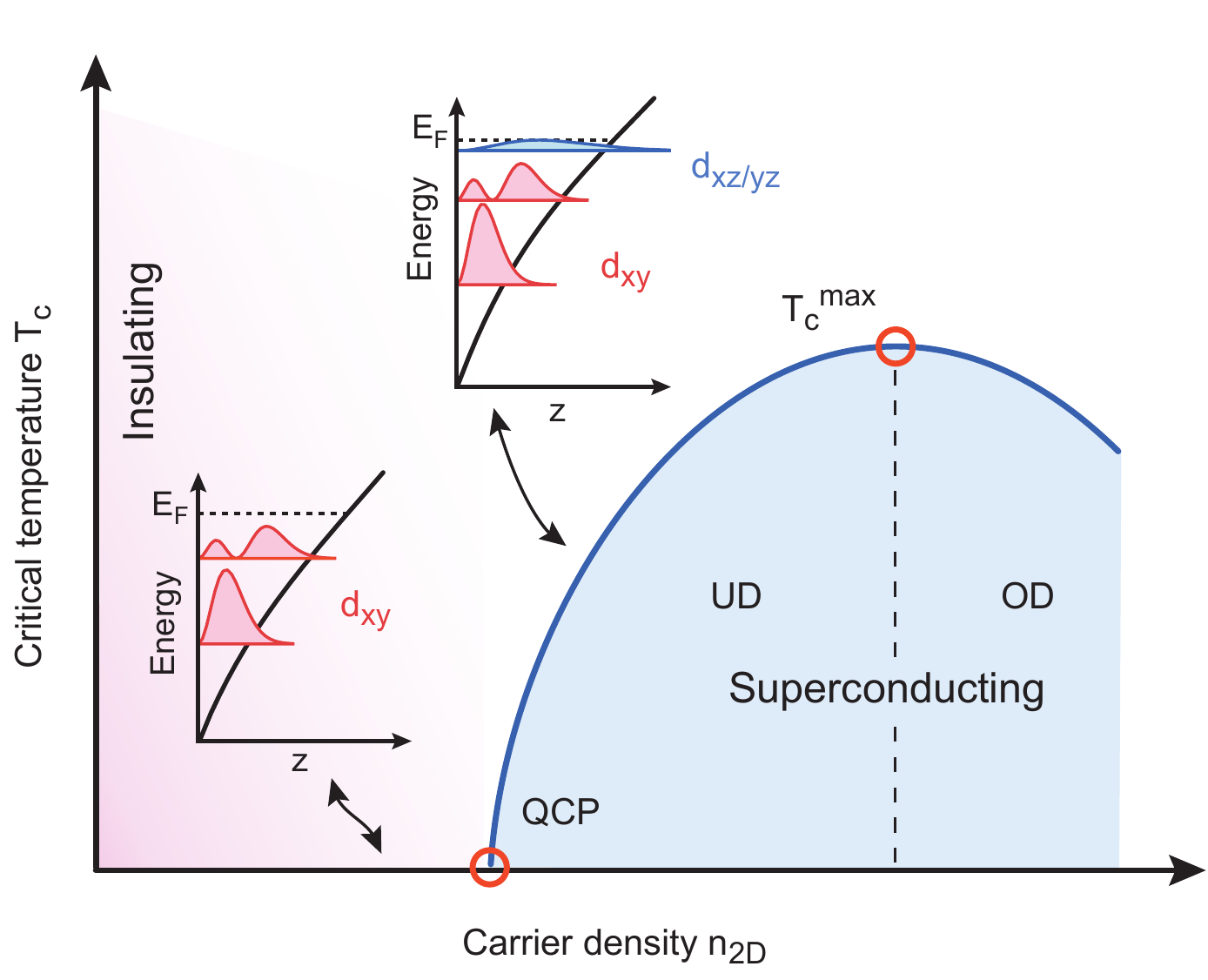}
\caption{Schematic description of the dome-shaped superconducting phase diagram of (001)-oriented \LAO/\STO interfaces.  The QCP marks the emergence of superconductivity at a critical carrier density corresponding the filling of the $d_\mathrm{xz/yz}$ heavy bands depicted by the sketch of the interfacial band structure. An optimal doping point corresponding to $T_c^\mathrm{max}$ separates an underdoped (UD) region where $T_c$ increases with $n_{\mathrm{2D}}$ and an overdoped (OD) region where $T_c$ decreases with $n_{\mathrm{2D}}$.  }
\end{figure}

\indent In this letter, we predict through self-consistent Schr\"odinger-Poisson equations that we can access different superconducting regimes related to different subband occupations close to the optimal doping levels. The suppression of $T_c$ in the overdoped regime can be delayed by adding electrons into the already populated $d_\mathrm{xz/yz}$ band with a top-gate. In turn, the action of a back-gate is associated with filling an additional high-energy $d_\mathrm{xy}$ subband, prospectively leading to the formation of a fragile s$_\pm$-wave superconducting state. The analysis of the superconducting properties of a double-gate  \LAO/\STO field-effect device evidences a bifurcation in the dependence of $T_c$ on $n_\mathrm{2D}$ that supports this scenario.\\ 

In multiband superconductors, a weakening of superconductivity has been predicted in the presence of strong disorder when the order parameters associated with each superconducting condensate have opposite signs because of a repulsive coupling. Although nodeless single-band superconductivity is essentially insensitive to scattering, in two-band s$_\pm$-wave superconductors, scattering processes between bands with opposite-sign gaps are pair-breaking, leading to a weakening of superconductivity \cite{trevisan,kogan2}. Following the approach of Trevisan et al. \cite{trevisan}, we interpret the optimal doping point $T_c^\mathrm{max}$ in \LAO/\STO as the filling threshold of a new band, which accommodates a second superconducting gap repulsively coupled to the first one. To substantiate this claim, in the following we examine the band structure in the interfacial quantum well by solving the coupled Schr\"odinger and Poisson equations self-consistently in the presence of a back-gate ($V_{\mathrm{BG}}$) and a top-gate ($V_{\mathrm{TG}}$). Before that, it is useful to have a look at the main difference between these two types of doping execution.  Schemes in Figure 2 provide an intuitive picture of the two situations. On the one hand, doping with $V_{\mathrm{BG}}$ repels the electrons from the interface by "pulling down" the conduction band in the \STO substrate, thus deconfining the 2-DEG. On the other hand, increasing  $V_{\mathrm{TG}}$ makes the confining potential sharper because of charges accumulation, which tends to attract the electrons towards the interface. \\ 

\begin{figure}[tb]
\includegraphics[width=8.5cm]{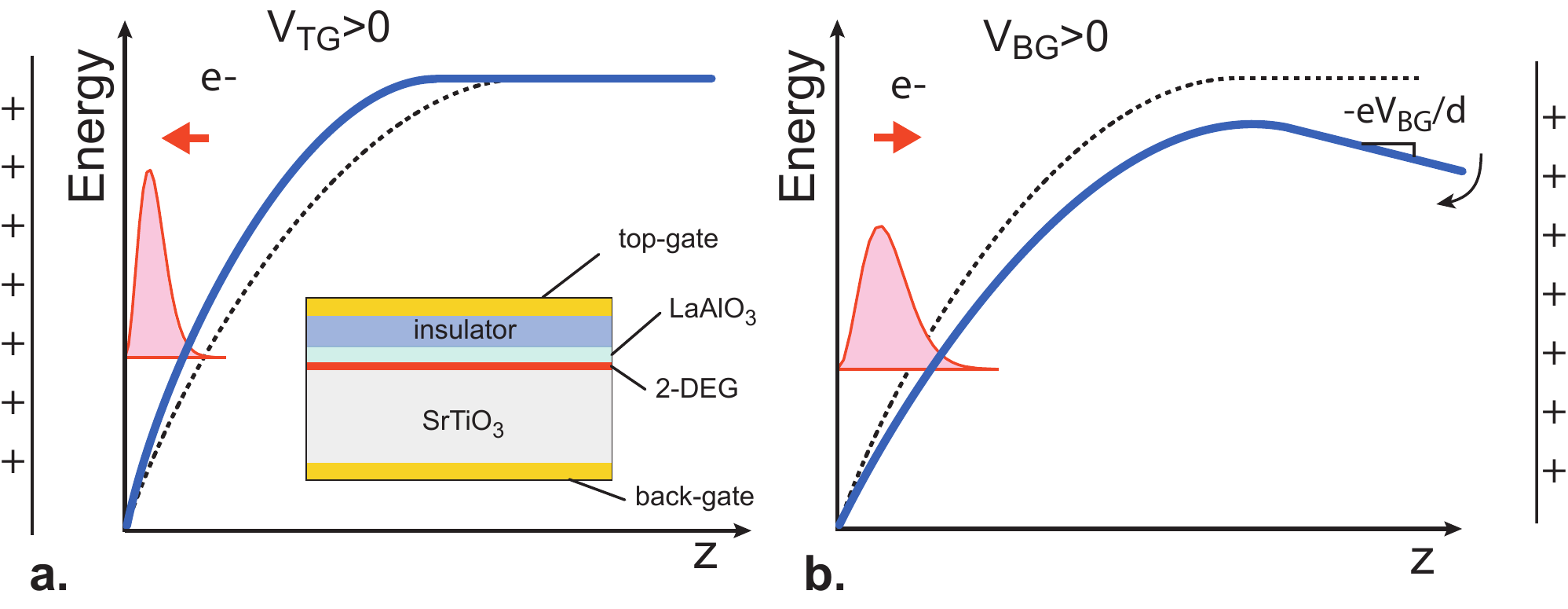}
\caption{Illustration of the difference between top-gating (panel a) and back-gating (panel b) and on the extension of the 2-DEG. While a positive top-gate marginally affects the potential well (dotted line: conduction band without an applied gate-voltage) and results in a further spatial confinement of the electronic wave packet (in red), a positive back-gate voltage causes a significantly larger band bending and tends to further delocalize electrons in \STO.  Inset) general scheme of the field-effect device considered in this study.} 
\end{figure}

\indent The numerical calculations, whose details are given in the Supplementary Information, account for the electric field dependence of the \STO permittivity, $\varepsilon_R$ \cite{NEVILLE:1972p3397} and the continuity of the potential in the whole substrate.  Figure 3 shows the results for a sequence of different back-gate and top-gate steps corresponding to carrier densities measured in a physical device discussed later.  In each panel, we plot the spatial dependence of the confining potential well, the energies, and the wave functions of the different $t_{2g}$ subbands. The insets illustrate the corresponding doping point (red circle) in the generic phase diagram of (001)-oriented \LAO/\STO interfaces. Starting with a carrier density of $n_{\mathrm{2D}}$ $\simeq$ 1.5 $\times$ 10$^{13}$ cm$^{-2}$  ($V_{\mathrm{BG}}=-20\mathrm{V},V_{\mathrm{TG}}=0\mathrm{V}$)  in the weakly insulating region, we see that three low-energy $d_\mathrm{xy}$ subbands are occupied (point \textbf{A} in panel 3$\textbf{a}$). When the electron density is further increased with the back-gate voltage, the $d_\mathrm{xz/yz}$ subbands start to be populated, leading to  a progressive delocalization of the 2-DEG in the \STO substrate upon gating. Because the $d_\mathrm{xz/yz}$ subbands have a  higher density of states than the $d_\mathrm{xy}$  one (by a factor $\simeq$ 4.4), superconductivity emerges as expected in a BCS scenario, and the $T_c$ rises with the back-gate voltage until it eventually reaches its maximum value (point \textbf{B} in panel 3$\textbf{b}$).  We now consider panels $\textbf{c}$ and $\textbf{d}$ and distinguish between a top-gate voltage step, $\Delta V_{\mathrm{TG}}$ = 50 V,  and a back-gate voltage step, $\Delta V_{\mathrm{BG}}$ = 5V, which both produce a similar carrier density variation $\Delta n_{\mathrm{2D}}$ $\simeq$ 4 $\times$ 10$^{12}$ cm$^{-2}$. For $\Delta V_{\mathrm{BG}}>0$, a new $d_\mathrm{xy}$ subband starts to be populated under the combined effects of the electron density increase and the deconfinement of the quantum well (panel 3$\textbf{c}$).  In contrast to the low-energy $d_\mathrm{xy}$ subbands that reside at the bottom of the quantum well, this new band extends deeper into the substrate. Benefiting from the coupling with the $d_\mathrm{xz/yz}$ band, a second superconducting gap opens in this band, and a s$_\pm$-wave superconducting state should takes place as proposed by Trevisan \textit{et al.} \cite{trevisan}. We therefore expect $T_c$ to decrease in the overdoped region because of interband scattering (point \textbf{C}). In contrast, for $\Delta V_{\mathrm{TG}}>0$,  the confining potential well becomes sharper as suggested in Fig. 2a, and the $d_\mathrm{xy}$ subband is repelled to higher energy (panel 3$\textbf{d}$).  The electron density in the $d_\mathrm{xz/yz}$ subbands increases, which should produce a further increase in $T_c$ in the single-gap superconducting regime (point \textbf{D}). Filling the high-energy $d_\mathrm{xy}$  subband is delayed, but it eventually occurs during a further increase of the top-gate voltage. The main result of these simulations, is the prediction of a bifurcation in the dependence of $T_c$ on $n_\mathrm{2D}$ upon top-gating or back-gating, that should occur in the vicinity of optimal doping. 
In order to test this scenario, we have studied the transport properties of a top-gated \LAO/\STO field-effect device.\\

\begin{figure}[tb]
\includegraphics[width=8.5cm]{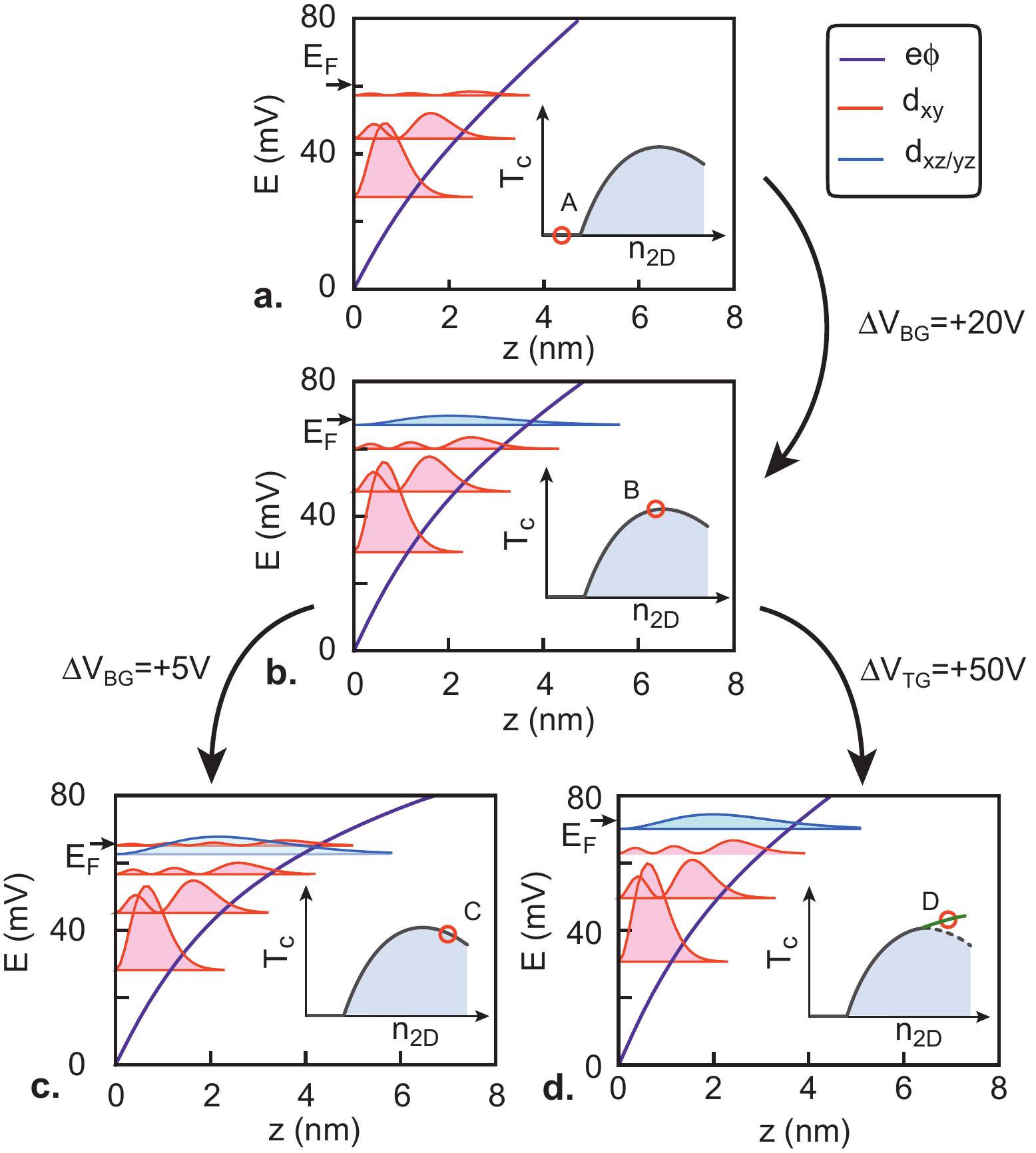}
\caption{Results of the numerical simulations of coupled Schr\"odinger and Poisson equations showing the spatial dependence of the confining potential well, the energies and the wave functions of the $d_\mathrm{xy}$ and $d_\mathrm{xz/yz}$ subbands for different carrier densities corresponding to a sequence of  back-gate and top-gate steps : [$V_{\mathrm{BG}}$ = -20V,$V_{\mathrm{TG}}$=0V] and $n_{\mathrm{2D}}$=1.5$\times$10$^{13}$cm$^{-2}$ (panel \textbf{a}),  [$V_{\mathrm{BG}}$ = 0V,$V_{\mathrm{TG}}$ = 0V] and $n_{\mathrm{2D}} = 2.75\times$10$^{13}$cm$^{-2}$ (panel \textbf{b}), [$V_{\mathrm{BG}}$ = +5V,$V_{\mathrm{TG}}$ = 0V] and $n_{\mathrm{2D}} = 3.16 \times$10$^{13}$cm$^{-2}$) (panel \textbf{c}), [$V_{\mathrm{BG}}$ = 0V,$V_{\mathrm{TG}}$ = +50V] and $n_{\mathrm{2D}} = 3.13 \times$10$^{13}$cm$^{-2}$ (panel \textbf{d}). The insets schematically indicate the corresponding carrier densities in the superconducting phase diagram (red circles).}  
\end{figure}

\indent  Whereas back-gate control of the 2-DEG properties is routinely realized in \STO-based interfaces, efficient top-gating has proven to be more challenging and has only been achieved in a limited number of studies. Nevertheless, the electrostatic control of the superconducting $T_c$ and the Rashba spin-orbit coupling has been demonstrated in field-effect devices with a top gate  evaporated either directly on the \LAO thin film \cite{hosoda,eerkes,forg,liu2015,goswami2015,chen}, or isolated by an additional dielectric layer \cite{hurand,hurand2,bal}. More recently, the manipulation of quantum orders at the mesoscopic scales with local top gates was demonstrated in Josephson Junctions \cite{thierschmann}, SQUIDs \cite{goswami2016}, quantum dots \cite{prawiroatmodjo} and quantum point contact devices \cite{jouan}. In this work, a 30$\times$10 $\mu m$ Hall bar was first fabricated in a \LAO(8 u.c)/\STO heterostructure by the amorphous \LAO template process \cite{stornaiuolo,hurand}.  After the deposition of a back-gate, a metallic top-gate separated by a \SN dielectric layer was also deposited on the Hall bar using a standard lithography and lift-off process.  More information on the fabrication of the device can be found in reference \cite{hurand}.  After the sample was cooled to 4K, both the top-gate voltage, $V_\mathrm{TG}$, and back gate voltage, $V_\mathrm{BG}$, were initially increased to their maximum positive value, beyond the saturation threshold of the resistance, to ensure that no hysteresis would occur during further gate sweeps \cite{biscaras3}.\\

 \begin{figure}[tb]
\includegraphics[width=8.5cm]{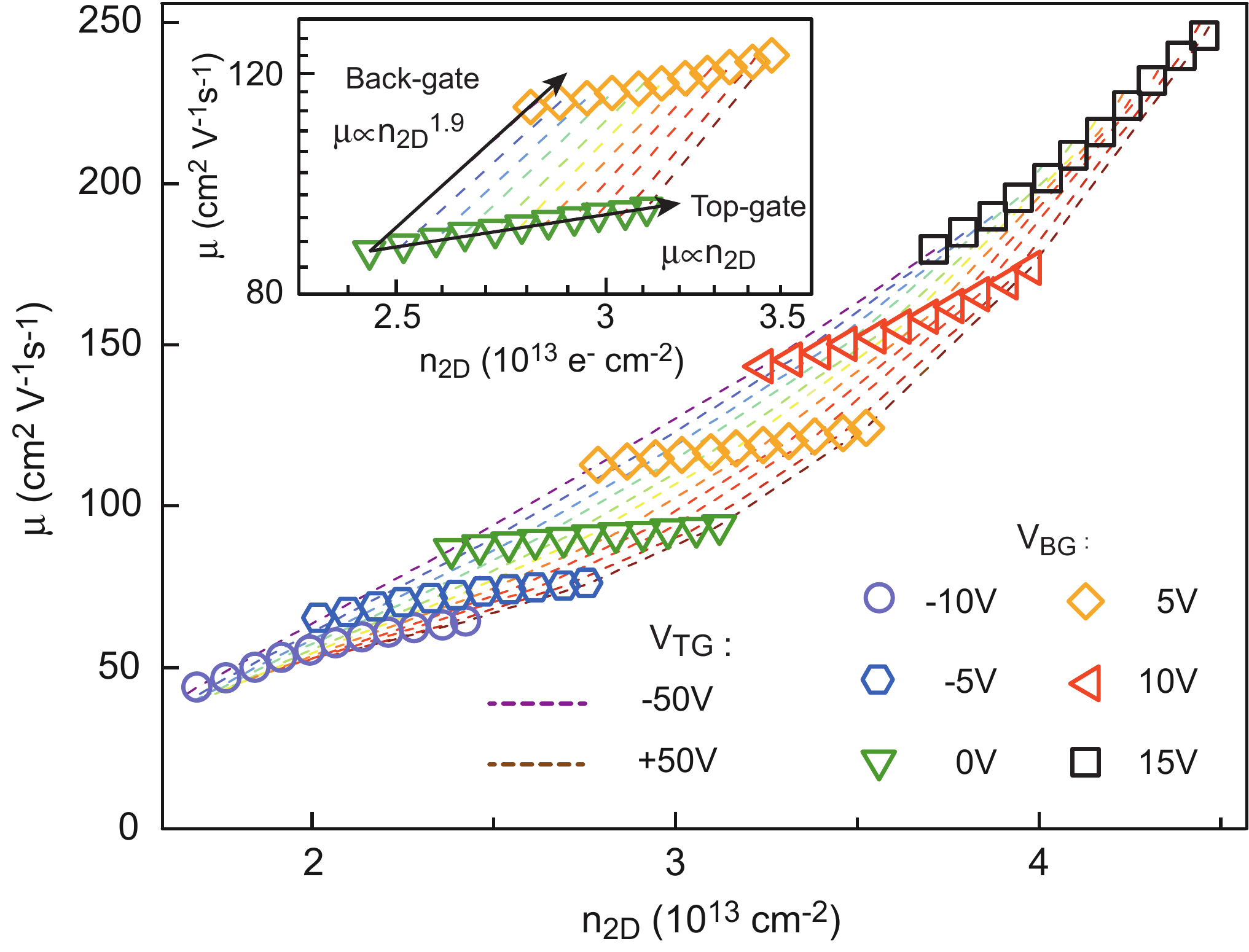}
\caption{Electronic mobility $\mu$, plotted as a function of  $n_{\mathrm{2D}}$. Symbols of a given color correspond to the same value of $V_{\mathrm{BG}}$.  Values of $V_{\mathrm{TG}}$ are represented by doted lines of different colors from $V_{\mathrm{TG}}$=-50V to $V_{\mathrm{TG}}$=+50V in step of 10V. Inset : zoom on the data at  $V_{\mathrm{BG}}=0$V and $V_{\mathrm{BG}}=5$V plotted on a logarithmic scale. Empirically, we find $\mu \propto n_{\mathrm{2D}}^\gamma$ with $\gamma_{\mathrm{BG}}\simeq1.9>\gamma_{\mathrm{TG}}\simeq1.0$.}
\end{figure}

\indent To understand the role of the two gates and check the operation of the device, we first compare the evolution of the  electronic mobility with carrier density when either the top-gate or the back-gate voltage is changed.  The total carrier density, $n_\mathrm{2D}$, is first extracted by combining the Hall effect and gate capacitance measurements at 4.2K \cite{biscaras2,singhCR,SI}, from which the mobility $\mu=1/en_\mathrm{2D}R_\mathrm{s}$ is deduced ($R_\mathrm{s}$ is the sheet resistance). Note that the electronic mobility considered here is the weighted sum of the mobilities in each subband. The results are summarized in Fig. 4.  In both cases, $\mu$ increases monotonically with $n_\mathrm{2D}$ in the entire doping range, but the slope is much sharper for $V_{\mathrm{BG}}$ than for $V_{\mathrm{TG}}$. Such behavior is consistent with previous results reported in the literature on conventional semiconducting hetero-interfaces \cite{ando}. It can be qualitatively understood by considering the sketches presented in Fig. 2. Whereas increasing  $V_{\mathrm{TG}}$ tends to attract the electrons towards the interface, increasing  $V_{\mathrm{BG}}$ deconfines the electrons deeper in the \STO substate which is naturally less disordered than the interface \cite{biscaras2, liu2015}. From a more quantitative perspective, in a 2-DEG with several 2D subbands, the mobility is predicted to scale as a power-law of the density ($\mu \propto n_\mathrm{2D}^\gamma$) \cite{ando}. Hirakawa et al. demonstrated theoretically, and confirmed experimentally, that the exponent, $\gamma$, is larger when using a back-gate rather than a top-gate \cite{hirakawa}. In line with this, we found that at each doping point of the phase diagram ($V_{\mathrm{BG}},V_{\mathrm{TG}}$), the variation of $\mu$ with $n_\mathrm{2D}$ can be locally approximated  by a power-law (inset Fig. 4). Although the exponents $\gamma_{\mathrm{BG}}$ and $\gamma_{\mathrm{TG}}$ vary in the phase diagram, the hierarchy $\gamma_{\mathrm{BG}} > \gamma_{\mathrm{TG}}$ is always satisfied in agreement with the prediction \cite{hirakawa}. For example,  for $V_{\mathrm{BG}}$ = 0 V we find $\gamma_\mathrm{BG}$ $\simeq$ 1.9 $>$ $\gamma_\mathrm{TG}$ $\simeq$ 1 (see inset Fig. 4), which corresponds to exponent values comparable with those measured in GaAs/Al$_{x}$Ga$_{1-x}$As heterojunctions \cite{hirakawa}. In particular, values of $\gamma$ close to 1 have been associated to Coulomb scattering from ionized donors in the Al$_{x}$Ga$_{1-x}$As layer \cite{ando}.

\begin{figure}[tb]
\includegraphics[width=8.5cm]{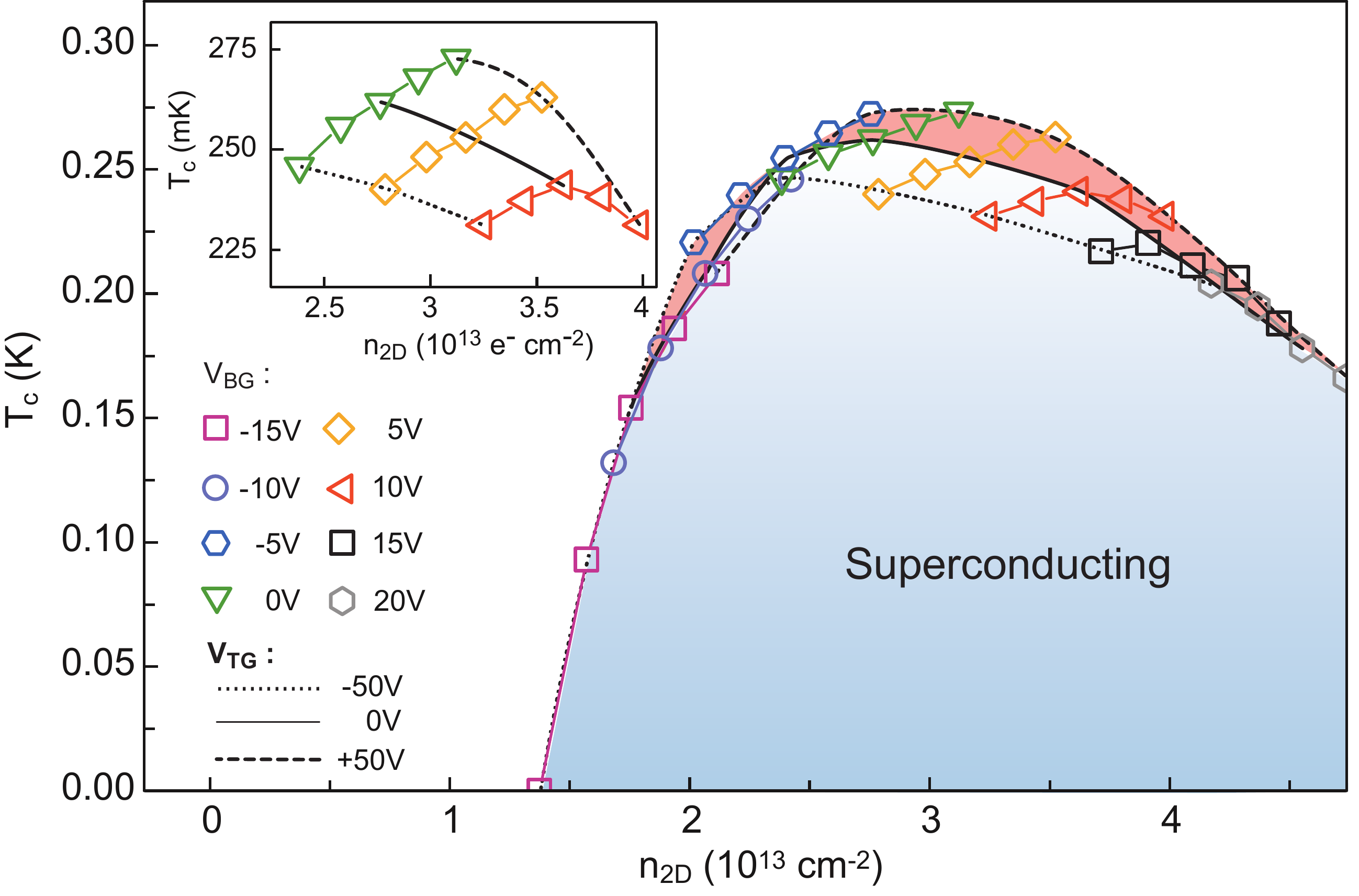}
\caption{Superconducting critical temperature defined at a 50$\%$ drop of the normal resistance plotted as a function of $n_{\mathrm{2D}}$, following the same color code as in Fig. 4. Resistive superconducting transition curves are shown in Supplementary Information \cite{SI}. Inset : magnification of the same data that emphasizes the bifurcation points in the dependence of $T_c$ on $n_\mathrm{2D}$. For example at the point ($V_{\mathrm{BG}}=0\mathrm{V},V_{\mathrm{TG}}=0\mathrm{V}$) a back-gate voltage step $\Delta V_{\mathrm{BG}}$ generates a decrease of $T_c$ whereas a top-gate voltage step $\Delta V_{\mathrm{TG}}$ generates an increase of $T_c$.}
\end{figure}

We now focus on the superconducting properties of our device. All the resistance vs temperature curves measured for the different top-gate and back-gate voltages are shown in the Supplementary Information. At low temperature, superconductivity emerges at a critical density, $n\simeq1.4\times 10^{13}$ $\mathrm{.cm}^{-2}$ (Fig. 5). The $T_{\mathrm{c}}$ then follows a dome-shaped dependence on $n_\mathrm{2D}$ with a maximum value of $T_{\mathrm{c}}^\mathrm{max}$ $\simeq$ 275 mK, similar to previous observations in  \STO-based interfaces \cite{caviglia}.  At the lowest and highest doping levels, $T_{\mathrm{c}}$ displays the same dependence on $n_\mathrm{2D}$ regardless of the gate geometry being used. However, our data reveal a very peculiar behavior  for intermediate doping close to the optimal doping : as predicted in our numerical simulations, a bifurcation is observed in the $T_c$ dependence on $n_\mathrm{2D}$ depending on which gate is used. For instance, at $V_{\mathrm{BG}}$ = 0V, doping with a back-gate reduces $T_{\mathrm{c}}$, whereas doping with a top-gate increases $T_{\mathrm{c}}$ (Fig 5). This corresponds to the situation described in panels \textbf{c} and \textbf{d} of Fig. 3 reporting the results of the numerical simulations incorporating the corresponding experimental carrier densities.  The region of higher $T_c$ in red color in the phase diagram of Fig. 5 is not accessible resorting solely to a back-gate. This observation supports the scenario in which the origin of the optimal doping point in the superconducting phase diagram is ascribed to the formation of a pair-breaking s$_\pm$-wave superconducting state, involving the $d_\mathrm{xz/yz}$ band and a  high-energy $d_\mathrm{xy}$ replica subband.\\

Other scenarios, mostly involving multiband effects, can be considered to explain the dome-like shape of $T_c$ as a function of gate voltage. For instance, Gariglio \textit{et al.} correlate the non-monotonic gate-dependent  $T_c$ to a non-monotonic variation of the three-dimensional  carrier density, $n_\mathrm{3D}$, at the interface \cite{gariglio}. The effective thickness of the 2-DEG, needed to determine $n_\mathrm{3D}$, is inferred from a systematic comparison of the parallel and perpendicular depairing magnetic fields in the superconducting phase diagram. A strong deconfinement of the 2-DEG with back-gate voltage takes place in the overdoped regime leading to a decrease in $n_\mathrm{3D}$ density while $n_\mathrm{2D}$ continues to increase. Although we do observe and increase of the 2-DEG spatial extension in our simulations, it does not seem to be sufficient to produce a drop in the $n_\mathrm{3D}$, whose gate evolution remains monotonic. Maniv \textit{et al.}  probed the area of the Fermi surface by using the Shubnikov-de Haas (SdH) effect and found that the population of mobile electrons associated with the highest energy occupied band, varies non-monotonically with gate voltage, thus explaining the gate dependence of $T_c$ \cite{maniv}. They ascribed this peculiar carrier density evolution to repulsive electronic correlations between bands that repels the highest energy band. We could not access the (SdH) regime in this work, but the analysis of non-linear Hall effect and capacitance measurements is consistent with a monotonous increase of both electrons populations in our case. More recently, an extended s-wave symmetry of the gap has been proposed to explain the gate dependence of $T_c$ \cite{zegrodnik}. Although little is known on the exact symmetry of the superconducting gap, tunneling and microwave conductivity experiments support  a nodeless isotropic gap \cite{richter,bert2,thiemann}.\\

In conclusion, we measured the low-temperature transport behavior of a field-effect \LAO/\STO device, whose electron density can be tuned simultaneously by means of a back-gate and a top-gate.  In the superconducting state, we evidenced a bifurcation in the $T_c$ dependence on $n_\mathrm{2D}$ :  close to the optimal doping point $T_c^\mathrm{max}$, a back-gate step generates a decrease in $T_c$, whereas a top-gate step produces an increase in $T_c$.  
We relate this behavior to filling a high-energy $d_\mathrm{xy}$  subband replica that leads to the formation of a two-gap s$\pm$-wave superconducting state involving the $d_\mathrm{xz/yz}$ and $d_\mathrm{xy}$ subbands, using self-consistent Schr\"odinger-Poisson calculation of the interfacial band structure. Pair-breaking inter-band scattering weakens the superconductivity in the overdoped regime hence providing a generic explanation for the the dome-shaped phase diagram of $T_c$.  Because both the density of states and the superfluid density  are expected to be weak for the $d_\mathrm{xy}$ subband,  the presence of two gaps may not be easily reflected in either electron tunnelling spectra or in superfluid stiffness. However, such suppression of $T_c$ in a two-gap superconducting state has been recently reported in (110)-oriented \LAO/\STO interface, a similar multiband system, albeit one having notable differences in the band structure and transport properties \cite{singhNM}. \\

\paragraph*{Acknowledgements}

The authors thank G. Venditti, M. Grilli and S. Caprara for very fruitful discussions. This work was supported by the french CNRS through a PICS program (S2S) and by ANR PRC (QUANTOP) and  by the French RENATECH network (French national
nanofabrication platform).The authors acknowledge received funding from the project Quantox of QuantERA ERA-NET Cofund in Quantum Technologies (Grant Agreement N. 731473) implemented within the European Union's Horizon 2020 Program and from the COST project Nanocohybri-Action CA16218.\\

\large\textbf{Author contributions}\\
\normalsize J.L. and N.B. supervised the project. E.L. fabricated the \LAO/\STO heterostructures by PLD under the supervision of A.B. and M.B. S.H. and A.J. made the top-gate devices with the help of C.F.-P. and  C.U..  S.H., AJ and G.S performed the measurements and A. J.  performed the numerical simulations. All authors contributed to the interpretation of the results. All the authors contributed to discussions of the results and writing of the manuscript.\\

\large\textbf{Competing interests:} \\
\normalsize The authors declare no competing financial interests.\\

\large\textbf{Additional information}\\
\normalsize\textbf{Supplementary Information} accompanies this paper. \\
\normalsize\textbf{Correspondence and requests} for materials should be addressed to A. J. and N. B.\\
\large\textbf{Data availability}\\
\normalsize All data that support the findings of this study are available from the corresponding authors upon reasonable request.\\

\newpage

\onecolumngrid
\section{Supplementary Information : Origin of the dome-shaped superconducting phase diagram in \STO-based interfaces}

\noindent\large{\textbf{1. Extraction of total carrier density}\\

\normalsize
We measured the Hall effect in a low magnetic field range ($B<5\mathrm{T}$) for different values of the back-gate voltage $V_\mathrm{BG}$ and top-gate voltage $V_\mathrm{TG}$.  As already reported  in \LAO/\STO 2-DEG, the Hall voltage is linear in magnetic field in the low-doping regime ($V_\mathrm{G}$ $<$  0), and the carrier density is correctly extracted from the slope of the Hall voltage V$_\mathrm{H}$  (i.e. n$_\mathrm{Hall}$= IB/eV$_\mathrm{H}$ where $I$ is the bias current  and $B$ the magnetic field). This is no longer the case in the high-doping regime ($V_\mathrm{G}>0$), where V$_\mathrm{H}$ is not linear with $B$ because of multiband transport \cite{biscaras2,singhCR}. In this case, n$_\mathrm{Hall}$ measured in the $B\rightarrow 0$ limit doesn't give the correct carrier density and show a non-meaningful decrease with gate voltage. The correct dependence of the total carrier density $n_\mathrm{2D}$ with $V_{\mathrm{BG}}$ can be retrieved from the charging curve of the gate capacitance $C(V_{\mathrm{BG}})$:

\begin{eqnarray} 
n_\mathrm{2D}(V_{\mathrm{BG}})=n_\mathrm{S}(V_{\mathrm{BG}}=-15 \mathrm{V})+\frac{1}{eA}\int_{-15\mathrm{V}}^{V_{\mathrm{BG}}}C(V)\mathrm{d}V
\end{eqnarray}  

where $A$ is the area of the capacitor.  The variation of $n_\mathrm{2D}$ with $V_{\mathrm{BG}}$ is not linear because of the field dependent dielectric permittivity of the \STO. On the other hand, for each value of $V_{\mathrm{BG}}$, $n_\mathrm{2D}$ varies linearly with $V_{\mathrm{TG}}$ as expected since \SN is a regular dielectric material with a field-independent dielectric constant.\\

\begin{figure}[h!]
\includegraphics[width=9cm]{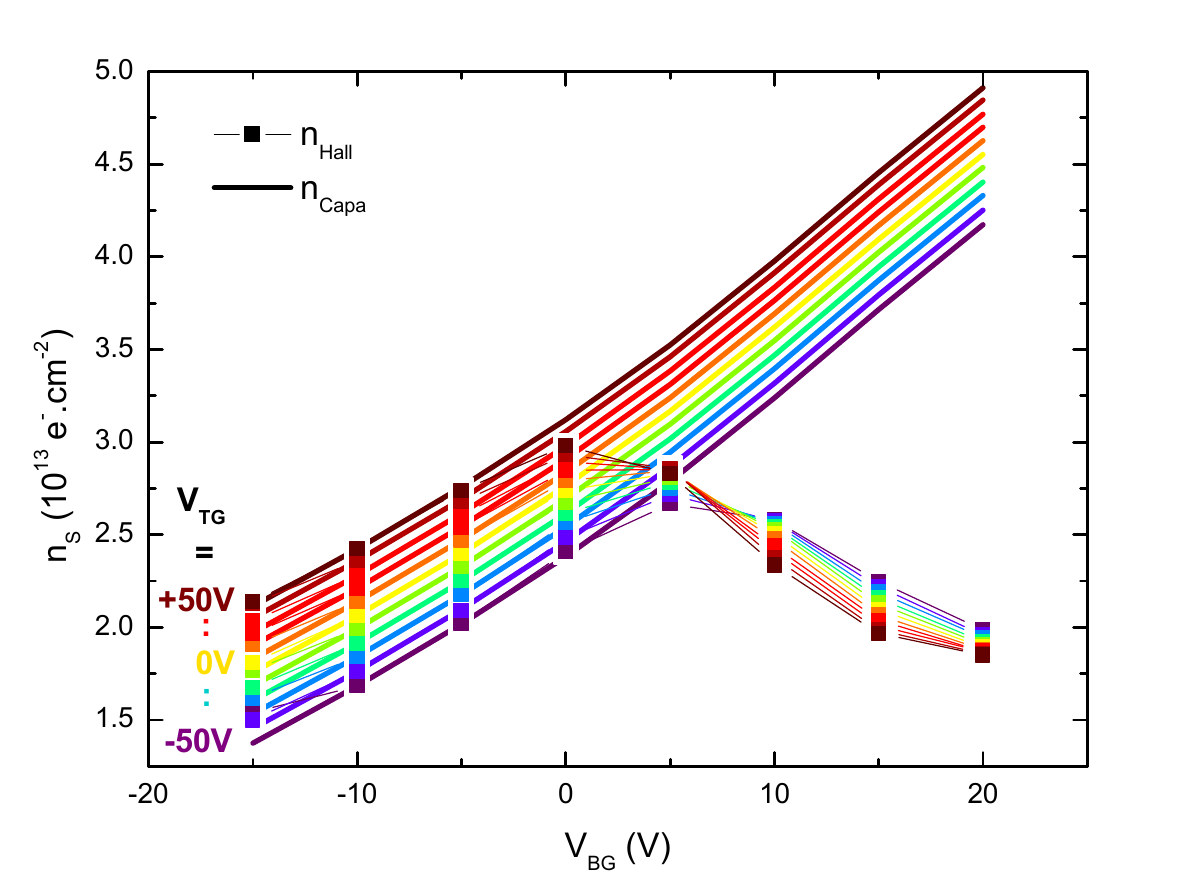}
\caption{Evolution of carrier density with top-gate and back-gate voltages.}
\end{figure}

\newpage

\noindent\large{\textbf{2. Superconducting transition}\\

\normalsize

Figure \ref{supra} shows the R(T) curves of the field-effect device for the different back-gate and top-gate voltages.\\
\begin{figure}[h!]
\includegraphics[width=12cm]{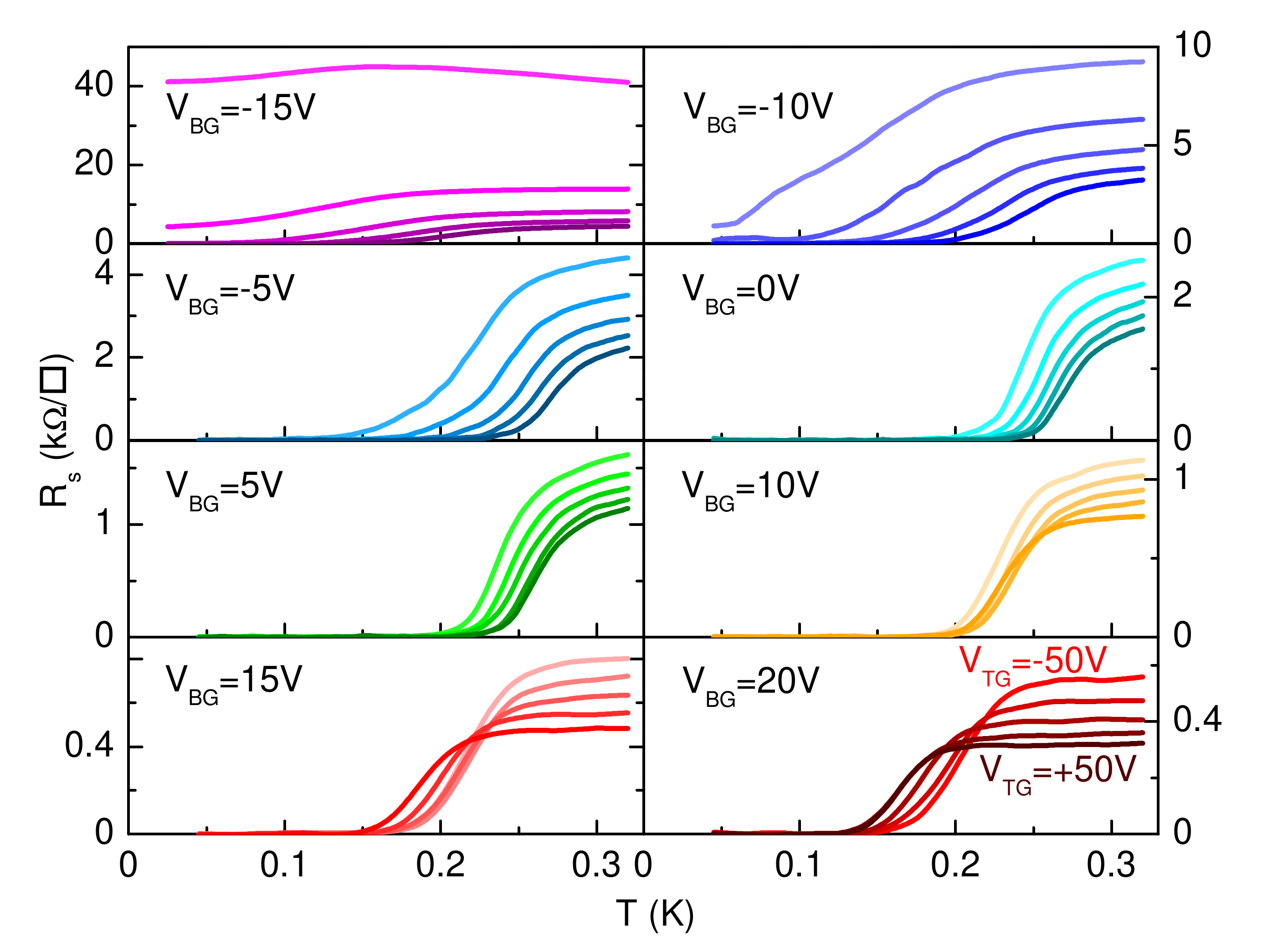}
\caption{Resistive superconducting transition for different values of top and back-gate voltages.}
\label{supra}
\end{figure}

\noindent\large{\textbf{3. Schr\"odinger-Poisson simulations}}\\
\normalsize

In the following we consider a double-gate field effect device as described in reference \cite{hurand}. Coupled Schr\"odinger and Poisson equations are solved numerically at the interface as proposed in reference \cite{stern}, in the effective mass approximation, with quantized electronic sub-bands, taking into account the non-linearity of \STO's dielectric permittivity. The \LAO side of the interface is modeled as an infinite barrier, since the electron gas is on the \STO side. We consider two type of parabolic band, $3d_{xy}$ ones with a in-plane effective mass taken as $m_{\parallel}^{xy}=0.7 \times m_0$ and a confinement mass of $m_z=14 \times m_0$ and the degenerate $3d_{xz/yz}$ band with $m_{\parallel}^{xz/yz}=3.13 \times m_0$ and $m_z=0.7 \times m_0$ \cite{Santander}. A two-dimensional electron gas confined in the z direction by a potential $e\phi(z)$, can be described by the following set of Schr\"odinger equations :

\begin{align}
\frac{d^2\psi_{xy}}{dz^2}+\frac{2m_z^{xy}}{\hbar^2}\left[E^{xy}+e\phi(z)\right]\psi_{xy}(z)&=0\\
\frac{d^2\psi_{xz/yz}}{dz^2}+\frac{2m_z^{xz/yz}}{\hbar^2}\left[E^{xz/yz}+e\phi(z)\right]\psi_{xz/yz}(z)&=0
\end{align}

where $\psi_{xy}(z)$ and $\psi_{xz/yz}(z)$ are the envelope wave functions for the $d_{xy}$ bands and $d_{xz/yz}$ bands respectively.\\ 

The boundary condition of the Maxwell-Gauss equation at the back side of the \STO substrate is imposed by the back gate voltage. We have therefore a continuous solution for the potential within the whole substrate which is realistic and essential to catch the physics of the system.\\

Since the regular Poisson's equation is not suited for spatially varying $\varepsilon_r$, the Maxwell-Gauss equation (which is it's parent equation) was solved numerically using 4th order Runge-Kutta method :

\begin{align}
&\nabla(\varepsilon_0\varepsilon_r(F(z))F(z))=-n^{tot}_{3D}(z)\label{eqpoisson1}
\\
&\varepsilon_r(F) = \varepsilon_r(F=\infty) + \frac{1}{A+B |F|}\label{eqpoisson2}
\end{align}

where $\varepsilon_r$ is the electric field dependent dielectric permittivity of STO with $A=4.097\times 10^{-5}$ and $B=4.907\times 10^{-10}$ m.V$^{-1}$ \cite{Neville1972}. A typical electric field can be calculated $F_c=A/B\simeq 10^{5}$ V.m$^{-1}$. In absence of electric field the dielectric constant reaches the value of bulk \STO $\varepsilon_r(F=0)\simeq \frac{1}{A}\simeq 23 000$. Above $F_c$, the dielectric permittivity is suppressed. \\

In \LAO/\STO interfaces, a large portion of electrons has been inferred to be localized in trapped states in the neighborhood of the interface. This rigid distribution of charge must be taken into account in the calculation of the quantum well profile \cite{gariglio}.  Note  that the presence of localized charges is crucial to maintain the confinement of the 2-DEG, in particular when a positive back-gate voltage is applied. Indeed, in their absence, the solution of the Schr\"odinger-Poisson equation would generate an unstable solution where the electrons spread to $z=\infty$ (corresponding to a Fermi energy at the exact top of the quantum well).  In reference \cite{gariglio},  a constant density of charges localised over a distance of 100 nm was considered. Here, we used a distribution of the form $N_{trap}(z)\propto e^{-z/Z_N}$ with a typical length $Z_N \sim 15$ nm. The density of trapped charges $N_{trap} = 3.10^{13}$ e$^{-}$/cm$^{2}$ was determined through the forming step of the 2-DEG. When the system is doped by applying a first positive gate voltage, the Fermi energy rises until the electrons can thermally escape from the quantum well \cite{Biscaras2014}. We therefore assume that at maximum voltage (+20V) the Fermi energy lies just below the top of the quantum well. $N_{trap}$ is chosen such that the numerical simulations reproduce this situation. Note that the values of $N_{trap}$ and $Z_N$ can affect the filling thresholds of the different $t_{2g}$ subbands in the simulations. However, the generic scenario discussed in the article, including the delayed filling of the $d_{xy}$ subband with the top-gate voltage remains largely unaffected by the choice of these parameters. \\

Using this set of parameters, Schr\"odinger's envelope equation is solved numerically from the input conduction band profile ${E_{\mathrm{C(in)}}}$ to find both energy levels and wave functions. Energy levels are filled to a self consistent Fermi energy $E_{\mathrm{F}}$ to match the sheet electronic density measured experimentally. Maxwell-Gauss equation is then integrated with the computed electronic density profile to give the output conduction band profile ${E_{\mathrm{C(out)}}}$. Both equations are solved iteratively, with the conduction band of the \emph{n}th iteration ${E_{\mathrm{C(in)}}^{(n)}}$ being $(1-f){E_{\mathrm{C(in)}}^{(n-1)}} + f {E_{\mathrm{C(out)}}^{(n-1)}}$, starting from a trial profile ${E_{\mathrm{C(in)}}^{(0)}}$. A deceleration factor $f$ taken as 0.02 ensured smooth convergence generally without oscillations until the sum of the squared error between ${E_{\mathrm{C(in)}}^{(n)}}$ and ${E_{\mathrm{C(out)}}^{(n)}}$ was less than $10^{-9}$ $\mbox{eV}/\mbox{\AA}$. Results of the simulations are presented in Figure 3 of the main article.

\end{document}